\RequirePackage{fix-cm}
\documentclass[twocolumn]{svjour3}          
\smartqed  

\usepackage{graphicx}
\usepackage[outdir=./,suffix=]{epstopdf}
\usepackage{amsmath}

\usepackage{amsfonts}
\usepackage[round,comma,authoryear,sort&compress]{natbib}
\usepackage{soul}
\usepackage{color}
\usepackage{fdsymbol}
\usepackage{subcaption}
\usepackage{placeins}
\usepackage{bm}
\usepackage{mathtools,tikz,caption}
\DeclareRobustCommand\sampleline[1]{%
  \tikz\draw[#1] (0,0) (0,\the\dimexpr\fontdimen22\textfont2\relax)
  -- (2em,\the\dimexpr\fontdimen22\textfont2\relax);%
}
\DeclareRobustCommand\sampledashdot{%
  \tikz\draw[thick, dash] (0,0) (0,\the\dimexpr\fontdimen22\textfont2\relax)
  -- (2em,\the\dimexpr\fontdimen22\textfont2\relax);%
}

\newcommand{\green}{\color[rgb]{0.1059,0.6196,0.4667}}
\newcommand{\orange}{\color[rgb]{0.8510,0.3725,0.0078}}
\newcommand{\blue}{\color[rgb]{0.4588, 0.4392, 0.7020}}
\newcommand{\magenta}{\color[rgb]{0.9059, 0.1608, 0.5412}}

\begin{document}

\title{Bias in particle tracking acceleration measurement
}
\subtitle{}

\author{John M. Lawson \and
		Eberhard Bodenschatz \and
		Cristian C. Lalescu \and		
		Michael Wilczek
}

\institute{J. M. Lawson \at
           Max Planck Institute for Dynamics and Self-Organisation \\
           Am Fassberg 17, 37077 G{\"o}ttingen \\
           Tel.: +49-551-5176-319 \\
           Fax:  +49-551-5176-302 \\
           \email{john.lawson@ds.mpg.de}
}

\date{Received: date / Accepted: date}

\maketitle

\begin{abstract}
We investigate sources of error in acceleration statistics from Lagrangian Particle Tracking (LPT) data and demonstrate techniques to eliminate or minimise bias errors introduced during processing. 
Numerical simulations of particle tracking experiments in isotropic turbulence show that the main sources of bias error arise from noise due to position uncertainty and selection biases introduced during numerical differentiation. 
We outline the use of independent measurements and filtering schemes to eliminate these biases. 
Moreover, we test the validity of our approach in estimating the statistical moments and probability densities of the Lagrangian acceleration. 
Finally, we apply these techniques to experimental particle tracking data and demonstrate their validity in practice with comparisons to available data from literature. 
The general approach, which is not limited to acceleration statistics, can be applied with as few as two cameras and permits a substantial reduction in the spatial resolution and sampling rate required to adequately measure statistics of Lagrangian acceleration.

\end{abstract}

\section{Introduction}
\label{sec:introduction}

The past fifteen years have seen the advent of Lagrangian Particle Tracking (LPT) methods applied to experimental fluid mechanics.
In a typical LPT experiment, time-series recordings are made of the motion of tracer particles seeded in the flow of interest.
The particles are then optically tracked using standard computer vision techniques \citep{Malik1993,Maas1993,Dracos1996,Hoyer2005,Ouellette2006}.
Reconstructed particle trajectories may then be numerically differentiated to obtain flow properties, such as velocity and acceleration, sampled at the particle position \citep{LaPorta2001,Crawford2004,Mordant2004}.
Multiple cameras (typically three or four) can be used to extend the technique to make three-dimensional (3D) measurements \citep{Nishino1989}.
LPT therefore readily lends itself to the examination of complex, unsteady, 3D flow phenomena from the Lagrangian frame, making it a natural choice in the investigation of problems such as turbulent transport and mixing \citep{Yeung2002,Holzner2008,ToschiBodenschatz2009} and intermittency \citep{Voth1998,Voth2002,LaPorta2001}.
In particular, the Lagrangian acceleration plays a central role in phenomena including turbulent dispersion \citep{Borgas1991,Salazar2009} and rain initiation in clouds \citep{Falkovich2002,Shaw2003,Bewley2013}.

In contrast to its popular cousin, Particle Image Velocimetry, LPT measurements typically require a low seeding concentration in order to unambiguously track individual tracer particles \citep{Raffel2007}.
This has the disadvantage that only sparse spatial information is available about the flow field at any given moment.
On average, however, LPT can resolve profiles of mean flow quantities down to sub-pixel accuracy, since instantaneous flow properties may be localised to the position of individual particles \citep{Kaehler2012}.
This makes LPT an ideal tool for the measurement of velocity profiles and other average field quantities, e.g. mean velocity and Reynolds stress profiles in the near wall boundary layer \citep{Kaehler2012b}.
In this context, systematic errors in statistics due to noise and resolution effects become important, since they cannot be eliminated by taking more data.

The measurement of Lagrangian accelerations using LPT has historically been particularly challenging \citep{Voth1998,Mann1999,Voth2002,Mordant2004,Berg2009}.
Since the position signal must be twice differentiated in time, Lagrangian acceleration measurements are very sensitive to noise.
Minimising position errors beyond $\sim 0.05$ pixel accuracy is difficult, since it arises from a variety of sources within the measurement process, including pixelisation, particle image overlap, sensor readout noise and quantisation errors \citep{Ouellette2006}.
This has led to the use of very high optical magnification to increase the position accuracy \citep{LaPorta2001,Mordant2004}. 
Furthermore, very high sampling rates are required to resolve very rapid fluctuations, which occur on time-scales shorter than the Kolmogorov scale \citep{Voth1998,LaPorta2001}. 

A common approach to mitigating noise has been to significantly oversample the position signal and subsequently apply a low-pass, finite-impulse-response filter with a large support \citep{LaPorta2001,Voth2002,Mordant2004,Crawford2004,Xu2007}.
The large filter support is intended to reduce position uncertainty, whilst increasing the sampling rate is intended to maintain high temporal resolution.
Such oversampling can lead to requirements for sampling rates on the order of tens of kilohertz in laboratory scale experiments \citep{LaPorta2001,Xu2007}.
Furthermore, a selection bias may be introduced when filtering is implemented as a discrete convolution, which requires special treatment for the ends of tracks and interpolated data \citep{Mann1999,Voth2002,Crawford2004}.

Once the sampling rate is fixed, the choice of filter time-scale is a compromise: short time-scales may not sufficiently filter the noise, whilst long time-scales may remove meaningful signal. 
Ideally, the optimal filter scale will be chosen in a range where a small change makes no difference to the statistics. 
This is not achieved in practice: the experiments of \citet{Voth1998,Voth2002,Mordant2004,Berg2009} found no range of filter scales where the acceleration statistics are independent of scale. 

As an alternative to oversampling, biases in statistical moments and probability distributions can be mitigated by making use of simultaneous, independent measurements made from two separate detectors measuring the same quantity \citep{Crawford2004,Mordant2004}. 
An observation of the measurement noise can be made by taking the difference between a pair of simultaneous measurements of the same quantity. 
If the measurements are independent, the noise distribution may be inferred, which can be used to compensate statistical moments and  probability distributions of noisy data \citep{StefanskiCarroll1990}. 

Although numerical differentiation cannot be avoided with LPT data, the selection biases introduced whilst filtering can be mitigated. \citet{Gesemann2016} and \citet{Schanz2016} have recently popularised the use of penalised cubic B-splines for fitting noisy particle tracking data to smooth curves. The advantage of this method is that the resulting fit is twice continuously differentiable along its entire support and the curve is interpolated where data is missing. However, no systematic study has examined the influence of this approach upon acceleration statistics. 

Filtering and noise are not the only sources of systematic measurement error. 
Other sources include preferential concentration of tracers due to tracers not following the flow \citep{Bec2006,ToschiBodenschatz2009,Gibert2012,Monchaux2012}, which introduce a selection bias effect, finite size effects \citep{Voth2002}, which introduce a spatial filtering, and tracking errors \citep{Xu2008}, which can result in spurious, large accelerations when the tracking algorithm begins following the trajectory of a different tracer.
However, effective mitigations already exist for these effects: preferential concentration and finite size effects can be avoided by using smaller, density-matched tracers which more faithfully follow the flow and tracking errors can be mitigated by reducing the seeding density or increasing the measurement frequency.

In this paper, we present a suite of methods to correct Lagrangian measurements for biases introduced by noise and filtering effects. 
These are based on the use of smoothing splines and independent measurements to correct noise biases. 
The methods are tested using numerical simulations of LPT measurements and via application to real experimental data.
This allows us to realise a substantial reduction in the temporal and spatial resolution required to adequately measure the statistics of Lagrangian accelerations. 

The paper is organised as follows. 
In \S\ref{sec:methodology}, we outline a suite of techniques to correct LPT data for systematic measurement errors introduced by noise and filtering effects. 
Subsequently, \S\ref{sec:datasets} outlines the generation of the experimental, numerical simulation and synthetic particle tracking datasets used to test these methods. 
The bias correction methods are validated with synthetic particle tracking data in \S\ref{sec:results}.
We then apply and further validate these techniques with real experimental data in \S\ref{sec:exp-results}. 
Concluding remarks are provided in \S\ref{sec:conclusion}.

\section{Methodology}
\label{sec:methodology}

We now present a suite of methods which may be used to compensate or minimise systematic biases introduced by noise and filtering effects. We first describe how simultaneous measurements can be used to correct statistics of noisy data in \S\ref{sec:methodology:noise} and outline how such simultaneous measurements can be applied to LPT data in \S\ref{sec:methodology:independent}. We then recap the use of filtering methods in \S\ref{sec:methodology:interpolation} and identify the penalised smoothing spline as a means to reduce selection bias when filtering data. 

\subsection{Noise}
\label{sec:methodology:noise}

Simultaneous measurements are a powerful tool to eliminate biases introduced by noise. 
Suppose one has two simultaneous, independent measurements $a_1 = a + \gamma_1$ and $a_2 = a + \gamma_2$ of a quantity $a$ (say, acceleration) with noise $\gamma_1$ and $\gamma_2$. 
We can then define sums and differences of these quantities
\begin{equation}
\label{eqn:a-sum-dif}
\begin{aligned}
\overline{a} &= \frac{1}{2}(a_2 + a_1) = a + \overline{\gamma}\\
\overline{\gamma} &= \frac{1}{2}(\gamma_2 + \gamma_1) \\
\widehat{a}      &= \frac{1}{2}(a_2 - a_1) = \frac{1}{2}(\gamma_2 - \gamma_1) = \widehat{\gamma} 
\end{aligned}
\end{equation}
which are denoted with an overbar ($\overline{a}$) or hat ($\widehat{a}$) respectively. 
From this, the $p^\textrm{th}$ moment of $a$ is given by
\begin{equation}
\label{eqn:a-moments}
\langle a^p \rangle = 
\langle \overline{a}^p \rangle 
- \sum_{k=1}^{p/2} \binom{p}{2k} \langle a^{p-2k}\rangle \langle \widehat{a}^{2k} \rangle
\end{equation}
where we have assumed the independence of $a$, $\gamma_1$ and $\gamma_2$ and that $\gamma_i$ are symmetrically distributed. 
The key ingredients are that $\overline{\gamma}$ and $\widehat{\gamma}$ have the same distribution, are independent of $a$ and their odd moments vanish. 
The first three even moments of $a$ are given by
\begin{equation}
\label{eqn:a-moments-list}
\begin{aligned}
\langle a^2 \rangle &= \langle \overline{a}^2 \rangle - \langle \widehat{a}^2\rangle \\
\langle a^4 \rangle &= \langle \overline{a}^4 \rangle - 6\langle a^2 \rangle \langle \widehat{a}^2\rangle - \langle \widehat{a}^4 \rangle \\
\langle a^6 \rangle &= \langle \overline{a}^6 \rangle - 15\langle a^2 \rangle\langle \widehat{a}^4 \rangle - 15\langle a^4 \rangle\langle \widehat{a}^2 \rangle - \langle\widehat{a}^6\rangle .
\end{aligned}
\end{equation}

The insight here is that the measured distribution of $\widehat{a}$ constitutes an observation of the distribution of $\overline{\gamma}$, which can be used to compensate systematic errors in statistics of $\overline{a}$.
Despite the simplicity of this approach, it is rarely used within the particle tracking literature. 
We are only aware of the experiments of \citet{Voth2002} and later \citet{Crawford2004} and \citep{Mordant2004}, who used a unique setup of 1D silicon strip detector sensors to make pairs of independent, 2D position measurements, which were used to estimate the error in their acceleration variance measurements.

The magnitude of the noise can be described by the signal to noise ratio, defined as
\begin{equation}
\label{eqn:snr}
\textrm{SNR} = -10\log_{10}(\langle a_m^2 \rangle/\langle a^2 \rangle - 1)
\end{equation}
and is measured in decibels (dB). 
Here, $\langle a_m^2 \rangle$ represents some measured variance and we have assumed the measurement noise is independent. 
Ordinarily, one cannot directly measure this quantity, because the ground truth variance $\langle a^2 \rangle$ is not known. 
However, since (\ref{eqn:a-moments-list}) may be used to estimate the ground truth $\langle a^2 \rangle$, an estimate of the signal to noise ratio can be made.  

The noise correction procedure can be extended to the level of the PDF. Let us write the PDF of a random variable $a$ as $f_a$. Since $\overline{a}$ is the sum of the independent variables $a$ and $\overline{\gamma}$, its PDF $f_{\overline{a}} = f_a \star f_{\overline{\gamma}}$ is given as the convolution of $f_a$ with $f_{\overline{\gamma}} = f_{\widehat{\gamma}} = f_{\widehat{a}}$. 
Then, we can make use of deconvoluting kernel density estimators \citep{StefanskiCarroll1990,WangWang2011} to deconvolve $f_{\overline{a}}$ with $f_{\widehat{a}}$. 
This technique is well known within the statistics literature, but has surprisingly not been adopted by the particle tracking community, despite its obvious suitability.

The deconvolution method is more readily understood in terms of characteristic functions of random variables. The characteristic function of variable $a$ with sample space frequency $t$ is defined as $\phi_a(t) \equiv \langle \mathrm{e}^{\mathrm{i}ta} \rangle = \int \mathrm{d}a f_a(a) \mathrm{e}^{ita}$ and is the Fourier transform of $f_{a}$. Under the preceding statistical independence and symmetry assumptions, we have:
\begin{equation}
\label{eqn:a-cf}
\phi_{\overline{a}}(t) 
\equiv \langle \mathrm{e}^{\mathrm{i}t(a + \overline{\gamma})} \rangle 
= \langle \mathrm{e}^{\mathrm{i}ta} \rangle \langle \mathrm{e}^{\mathrm{i}t\overline{\gamma}} \rangle 
= \phi_{a}(t) \phi_{\widehat{a}}(t).
\end{equation}
This allows us to correct our measured distribution $f_{\overline{a}}$ using the distribution of the error difference $f_{\widehat{a}}$. Na{\"i}vely, we might calculate $f_a$ from $\phi_{a}(t) = \phi_{\overline{a}}/\phi_{\widehat{a}}$. However, the size of the compensation becomes quite large for large frequencies, so we introduce a kernel function $\phi_K(t)$ e.g. \citep{WangWang2011}
\begin{equation}
\label{eqn:phi_K}
\phi_K(t) = 
\left\{\begin{matrix}
(1 - t^2)^3 & -1 \le t \le 1
\\ 
0 & \textrm{otherwise}
\end{matrix}\right.
\end{equation}
with a characteristic bandwidth $h$ to define our deconvolution kernel $\phi_{L}(t)$
\begin{equation}
\label{eqn:cf-deconvolution}
\phi_{L}(t)  = \frac{\phi_K(t/h)}{\phi_{\widehat{a}}(t)}.
\end{equation}
The corrected PDF $\widetilde{f}_a$ can then be obtained from the inverse Fourier transform of $\widetilde{\phi}_{a}(t) = \phi_{\overline{a}}(t) \phi_{L}(t)$. 
The scale of the deconvolution kernel is given by the parameter $h$.  
The choice of kernel scale comes down to a compromise between bandwidth and statistical uncertainty. 
\citet{WangWang2011} provide a number of methods to choose the bandwidth; we also discuss the choice of bandwidth in \S\ref{sec:results:acceleration}. 

In practice, we obtain $\phi_{\overline{a}}$ and $\phi_{\widehat{a}}$ from the discrete Fourier-transform of fine-grained histograms of $\overline{a}$ and $\widehat{a}$. Subsequently, we obtain the histogram of $f_a$ from the inverse transform of (\ref{eqn:cf-deconvolution}).
Standardised PDFs are always scaled by $\sigma_a \equiv \langle a^2 \rangle^{1/2}$ with $\langle a^2 \rangle$ obtained from (\ref{eqn:a-moments-list}).

\subsection{Making Independent Measurements}
\label{sec:methodology:independent}

To use the noise correction techniques described in \S\ref{sec:methodology:noise}, we need to make simultaneous independent measurements of velocity and acceleration. 
In this section, we outline a novel approach to doing so, by making partial measurements of the velocity or acceleration from independent cameras. 

The image space velocity $\dot{\bm{y}_i}\in \mathbb{R}^2$ of a particle in camera $i$ is given by 
\begin{equation}
\label{eqn:img-jacobian}
\dot{\bm{y}_i} = \bm{J}_i(\bm{x}) \bm{u} 
\end{equation}
where $\bm{u} \in \mathbb{R}^3$ is the particle's velocity and $\bm{J}_i(\bm{x}) = \partial \bm{y}_i / \partial \bm{x}$ is the Jacobian of its projected position $\bm{y}_i \in \mathbb{R}^2$ in camera $i$ with respect to its position $\bm{x}\in\mathbb{R}^3$. 
Treating the projection as locally linear, one can also approximate the image space acceleration as $\ddot{\bm{y}}_i = \bm{J}_i\bm{a}$, where $\bm{a}\in \mathbb{R}^3$ is the particle acceleration.
Since these expressions have the same form, we will continue our analysis just for the velocity. 

The Jacobian has a null space vector $\bm{n}_i = \bm{n}_i(\bm{x})$. For a pinhole camera model, this is is parallel to the viewing direction. We can then solve (\ref{eqn:img-jacobian}) for $\tilde{\bm{u}}_i = \bm{u} - (\bm{u}\cdot\bm{n}_i) \bm{n}_i$ with the constraint $\tilde{\bm{u}}_i\cdot\bm{n}_i = 0$ to obtain a projection of the velocity from that measured in a single image. With this formulation, we see that a pair of cameras $i,j$ can make independent measurements of the same velocity component in the direction $\bm{n}_i \times \bm{n}_j$.

We can combine measurements from selected sets of cameras to measure all three components by solving (\ref{eqn:img-jacobian}) in the least-squares sense, as given by
\begin{equation}
\label{eqn:vel-lsq}
\bm{u} = \Big(\sum_i \bm{J}_i^T \bm{J}_i\Big)^{-1} \sum_i \bm{J}_i^T \dot{\bm{y}}_i.
\end{equation} 
In this way, we can make independent measurements of the velocity from different sets of cameras. For example, with three cameras, we could use one principal camera and a second pair to get independent measurements of two components simultaneously. With four cameras, simultaneous independent measurements of all three components can be made by using two pairs.

We note, in passing, that it may also be desirable to measure velocity increments $\Delta\bm{u} = \bm{u}(\bm{x}+\bm{r},t+\tau) - \bm{u}(\bm{x},t)$. With two or more cameras, this is no problem. With a single camera, technically only the component parallel to $\bm{n}_i(\bm{x}) \times \bm{n}_i(\bm{x}+\bm{r})$ may be measured. For the pinhole camera model, this is perpendicular to the separation vector $\bm{r}$, i.e. it corresponds to a transverse velocity increment. However, if the measurement volume is far from the camera pinhole, $\bm{n}_i(\bm{x})$ only varies slightly within the volume. In this case, two components of $\Delta\bm{u}$ can be measured, which are approximately perpendicular to $\bm{n}_i(\bm{x}+\bm{r}/2)$. 

The image space velocity $\bf{\dot{y}}_i$ can be obtained from numerical differentiation of the image space position.
Where necessary, it should be obtained from independent interpolations in image space to avoid introducing correlations in errors across cameras. 
The Jacobian can be readily computed from the camera model. 
For the purpose of illustration, we demonstrate this for the pinhole camera model used presently.
In this case, the transformation mapping $\bm{x}$ to $\bm{y}_i$ and its Jacobian $\bm{J}_i(\bm{x})$ can be written in the form \citep{Hartley2003}
\begin{equation}
\label{eqn:pinhole}
\bm{y}_i = \frac{\bm{T}_i(\bm{x}-\bm{x}_{c,i})}{\bm{\Lambda}_i(\bm{x} - \bm{x}_{c,i})}
~\textrm{and}~
\bm{J}_i = \frac{\bm{T}_i - \bm{y}_i\bm{\Lambda}_i}{\bm{\Lambda}_i(\bm{x}-\bm{x}_{c,i})}.
\end{equation}
Here, $\bm{x}_{c,i}$ is the location of the camera pinhole in object space, whilst $\bm{T}_i$ and $\bm{\Lambda}_i$ are $2\times 3$ and $1\times 3$ matrices parameterising the camera's orientation, magnification and distortion.

Our main assumption is that measurements of image space velocity and acceleration in separate cameras constitute independent measurements with independent errors. Peak-locking, i.e. systematic errors in the measurement of particle position, may violate this criterion since this effectively introduces a quantisation error into the velocity or acceleration. This assumption is also violated when using the Shake The Box technique \citep{Schanz2016}, since particle positions are jointly optimised across cameras. It may also break down when particles become very close to one another, since shadowing effects may become correlated across cameras. We also neglect the error in $\bm{J}(\bm{x})$ that contributes to error in $\bm{u}$. 

\subsection{Track Filtering and Interpolation}
\label{sec:methodology:interpolation}

The conventional approach to obtaining the velocity and acceleration of fluid tracers from particle tracks is to apply a finite-difference method in combination with some level of smoothing filter (e.g. \citet{Mann1999,Voth2002,Mordant2004,Crawford2004}). In our analysis, we pick a Gaussian-weighted least-squares approximation to the position, velocity and acceleration by convolving with a set of discrete filter kernels. For some filter of scale $w$ and support of $l$ samples, the kernels 
\begin{align}
\label{eqn:pos-filter}
h_n  &= C_1 \exp(-n^2/w^2) \\
\label{eqn:vel-filter}
h_n' &= n C_2  \exp(-n^2/w^2)
\end{align} and \begin{equation}
\label{eqn:acc-filter}
h_n'' = C_3 (2n^2/w^2-1) \exp(-n^2/w^2) - C_4
\end{equation}
can be convolved with the discretely sampled position signal $\mathbf{x}(t_n)$ to obtain the filtered position, velocity and acceleration, respectively. 
We choose a support of $l = 3w$ with $n \in {-l/2, ..., l/2}$. 
The coefficients $C_1 ... C_4$ are chosen to satisfy the normalisation conditions $h_n\star 1 = 1$, $h_n'\star 1 = 0$, $h_n''\star 1 = 0$, $h_n' \star (n\Delta t) = 1$ and $h_n''\star (n\Delta t)^2 = 2$, where $\Delta t$ is the sampling period. More details on the choice of this filter can be found in \citet{Mordant2004}. 

A limitation of this approach is that the filtered quantities are undefined near where data is missing (e.g. where tracks have been reconnected) and the ends of tracks, where samples are not available to apply the convolution kernel in (\ref{eqn:pos-filter}). As we show in \S\ref{sec:results:acceleration}, simply ignoring these data leads to a selection bias in the acceleration statistics. 

An alternative approach to filtering noisy data is to fit a smoothing spline \citep{EilersMarx1996,Gesemann2016}. 
The idea is to fit a spline curve $g(t)$ to the data $(t_i,y_i)$, $i = 1 ... m$ which makes a tradeoff between the closeness of the fit and the roughness of the curve. 
\citet{Gesemann2016} proposed the use of a cubic spline fit with a third-order roughness penalty, which minimises the following objective function:
\begin{equation}
\label{eqn:spline-o3-penalty}
\sum_{i=1}^{m} (g(t_i) - y_i)^2 + f_{s}\lambda\int_{t_1}^{t_m} g^{(3)}(x)^2 \mathrm{d}t.
\end{equation}
Here, $g^{(3)}(t)$ is the third derivative of the curve, $\lambda$ parameterises the level of smoothing and $f_s = 1/\Delta t$ is the sample rate. 
Numerically, we implement this by performing a penalised linear fit of a set of $m+2$ B-spline curves to the data. 
Details of how to do this can be found in \citet{EilersMarx1996}.

This approach has the advantage of generating a smooth representation of the underlying data which is continuous in its second derivative over the entire length of the data. As such, it may interpolate missing data in particle tracks, e.g. where tracks have been reconnected. Whilst end effects are present (the smoothing criterion results in $g^{(3)} = 0$ at the ends), this effect is much less severe than for a simple convolution filter, since the data at the ends are still represented. 

The filter parameter $\lambda$ determines the tradeoff between the smoothness and the closeness of the fit. When $\lambda = 0$, the chosen spline interpolates the data exactly. As $\lambda \rightarrow \infty$, the fit becomes a linear least squares regression to a quadratic polynomial. The frequency response approximates a sixth-order low-pass Butterworth filter with cutoff frequency $f_{c} = 1/(2\pi\lambda^{1/6})$. 

In order to make a fair comparison between filters, we compare their performance in terms of their Equivalent Noise Bandwidth (ENBW). The ENBW is the bandwidth of the equivalent brick-wall filter with the same integrated (white) noise power and may be calculated from the impulse response $h_n$ as \citep{Elliott1987}
\begin{equation}
\label{eqn:enbw}
f_{\textrm{ENBW}} = \frac{f_{s}}{2} \frac{\sum_n h_n^2}{(\sum_n h_n)^2}.
\end{equation}

\section{Experimental and Numerical Datasets}
\label{sec:datasets}

To test the methods described in \S\ref{sec:methodology}, we have conducted  laboratory particle tracking experiments and numerical simulations of particle tracking. 
Our numerical simulations consist of two aspects: direct numerical simulation of tracer particles in homogeneous isotropic turbulence and the subsequent simulation of experimental measurement of particle tracks. 
We first describe our experimental measurement, then the DNS and the synthetic particle tracking.

\subsection{Laboratory Experiment}
\label{sec:datasets:experiment}

\begin{figure}
\centering
\begin{subfigure}[h]{0.275\textwidth}
	\includegraphics[trim=0cm 0cm 0cm 0cm,clip,width=1\textwidth]{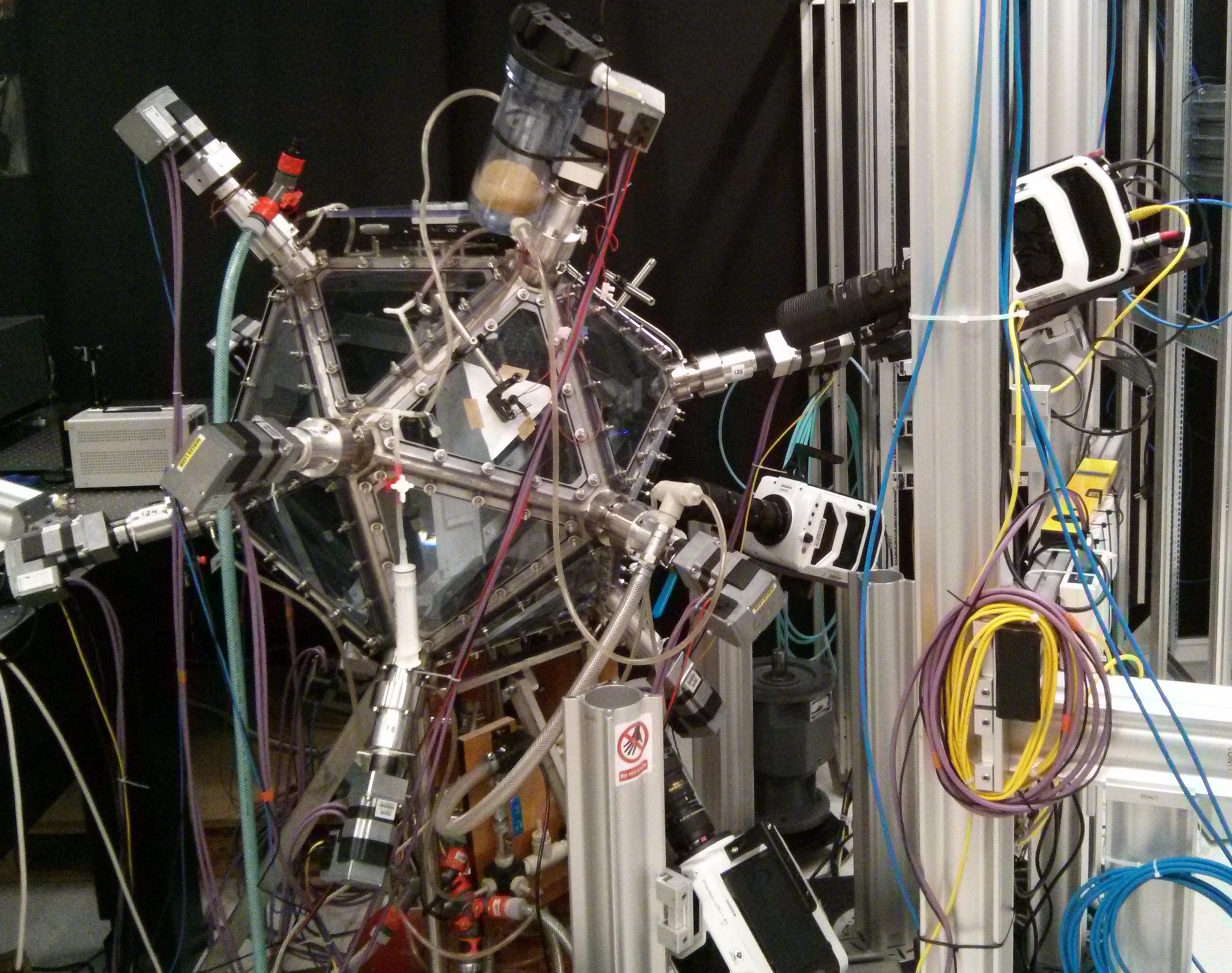}
	\caption{Experimental Setup}
	\label{fig:lem}
\end{subfigure}
\begin{subfigure}[h]{0.20\textwidth}
	\includegraphics[trim=0cm 0cm 0cm 0cm,clip,width=.87\textwidth]{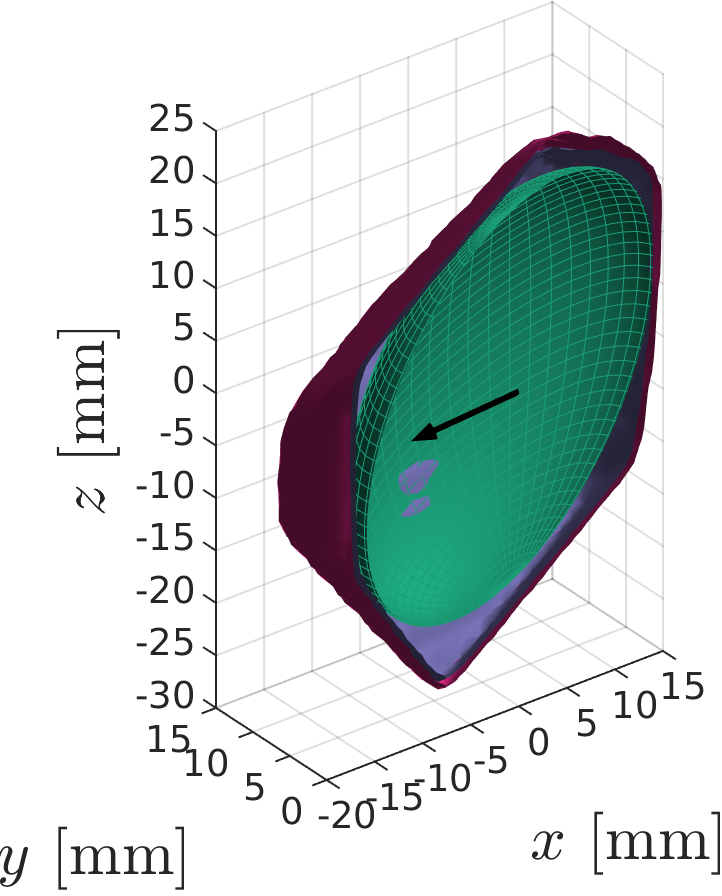}
	\caption{Measurement Volume}
	\label{fig:seeding-density}	
\end{subfigure}
\caption{(\subref{fig:lem}) Illustration of the Lagrangian Exploration Module and (\subref{fig:seeding-density}) a cross section of the experimental measurement volume. The magenta and blue isosurfaces of constant seeding density encapsulate $95\%$ and $99\%$ of the measured data, respectively. The acceleration of tracers is only sampled inside the green ellipsoidal volume. Pairs of independent measurements of the acceleration component are made in the direction shown by the black arrow.}
\label{fig:exp-setup}
\end{figure}

We conducted Lagrangian particle tracking experiments of low Stokes number tracer particles in deionised water in the homogeneous, isotropic turbulence generated in our Lagrangian Exploration Module (LEM) facility,  illustrated in Figure \ref{fig:lem}. We refer the reader to \citet{Zimmermann2010} for a full description of the facility and provide only a brief outline here. 

The LEM consists of an icosahedral tank with transparent polyacrylamide windows and impellers at each of its twelve vertices. This configuration allows us to generate a turbulent flow at the center of the tank which is statistically homogeneous and isotropic, with a mean flow speed below 10\% of the fluctuating velocity. The turbulence intensity and hence Reynolds number is adjusted by varying the rotation rate of the impellers. In the present experiments, we chose an isotropic forcing with all impellers rotating at the same frequency (between $60$ and $960\textrm{rpm}$). The temperature of the deionised water is maintained at $20^\circ$C by cooling plates at the top and bottom of the tank, whose cooling power is regulated by a closed-loop feedback controller. This maintains the kinematic viscosity of the water at $\nu = 1.004 \textrm{mm}^2/\textrm{s}$ and mass density at $\rho_f = 0.997 \textrm{kg}/\ell$.

Three high speed cameras (Phantom V2511, Vision Research) equipped with Nikon 200mm macro lenses and 2x teleconverters observe a measurement volume at the center of the LEM. The camera configuration and measurement volume are illustrated in Figure \ref{fig:exp-setup}. The region mutually visible across all cameras spans approximately $37\times 26\times 42\textrm{mm}$.
A 70W self-built pulsed Nd:YAG laser is coupled to beam forming optics to illuminate polystyrene tracer particles with diameter $d_p = 40\mu\textrm{m}$ and mass density $\rho_p = 1.05\textrm{kg}/\ell$ (TS40, Microbeads AS). For each experimental condition in Table \ref{tbl:datasets} we acquired 4000 independent time-series with O(1000) particles per image, corresponding to a seeding concentration around $47$ particles per $\textrm{cm}^3$. Each 3500 frame movie is downloaded over 10GBit Ethernet and saved in a sparse format by retaining only pixels and their neighbours with brightness above a specified threshold. Sparsification reduces the storage requirement by 90-95\% over uncompressed images. The fast download and sparsification process has allowed us to acquire very large datasets with O($10^{10}$) data-points per set. 

We present data collected at five different Reynolds numbers, detailed in Table \ref{tbl:datasets}. 
The integral length-scale $L_{int} = u'^3 / \epsilon \simeq 60\textrm{mm}$, defined in terms of the mean dissipation rate $\epsilon$ and the root-mean-square velocity fluctuation $u'$, is approximately independent of the impeller rotation rate $f_I$. 
The mean dissipation rate was obtained from measurements of the velocity-acceleration structure function \citep{Mann1999}.
The sampling rate $f_s$ is chosen to be 43-92 times faster than the Kolmogorov frequency, which is necessary to capture the heavy tails of the acceleration distribution \citep{Crawford2004,Mordant2004}. 
This results in a relatively small RMS inter-frame particle displacement of around 0.5 pixels.
The Stokes number $St \equiv \tau_p / \tau_\eta$ is a measure of the response time $\tau_p = d_p^2 / 12\beta\nu$ of our tracers in comparison to the Kolmogorov scale $\tau_\eta = (\nu/\epsilon)^{1/2}$, where $\beta = 3\rho_f/(2\rho_p+\rho_f)$. 
Since the Stokes' numbers achieved are small, we expect filtering effects due to particle size to be negligible \citep{LalescuWilczek2018b,Voth2002}.
However, we note that preferential concentration effects may influence the acceleration statistics at the highest Reynolds numbers \citep{Bec2006,Gibert2012}. 

\begin{table}[h]
\centering
\caption{Characterisation of experimental datasets. Details of the synthetic particle tracking simulation, detailed in \S\ref{sec:datasets:syn}, are also included.}
\label{tbl:datasets}
\setlength\tabcolsep{4.5pt}
\begin{tabular}{lllllllll}
\hline
$R_\lambda$ & $f_{I}$ & $u'$                     & $\epsilon$                   & $L_{int}$    & $\tau_\eta$   & $\textrm{St}$ & $f_{s}\tau_\eta$ & $\eta$        \\
-           & Hz      & $\textrm{mm}/\textrm{s}$ & $\textrm{cm}^2/\textrm{s}^3$ & $\textrm{mm}$ & $\textrm{ms}$ & -             & -                & $\textrm{px}$ \\ \hline
109         & 1       & 14.6                     & $0.568$                      & 55            & 133           & 0.001         & 92.3             & 11            \\
203         & 3       & 47.7                     & $18.8$                       & 58            & 23.1          & 0.006         & 64.2             & 4.6           \\
352         & 8       & 135                      & $402$                        & 61            & 5.00          & 0.013         & 62.5             & 2.1           \\
438         & 12      & 207                      & $1420$                       & 62            & 2.66          & 0.052         & 66.4             & 1.6           \\
504         & 16      & 276                      & $3390$                       & 62            & 1.72          & 0.080         & 43.0             & 1.3           \\ \hline
190         & -       & 46.9                     & $19.9$                       & 52  
& 22.4          & 0             & 98.1             & 4.5 \\ \hline
\end{tabular}
\end{table}

Particle tracking is performed using an in-house code which implements a version of the predictor-corrector tracking algorithm described in \citet{Ouellette2006}. We describe the procedure here only briefly. In each frame, particle images are identified using a local maximum criterion and their image centers obtained using a three-point Gaussian fit. Then, existing particle tracks are extrapolated using a quadratic polynomial fit and are associated with particle images if possible. Tracks are terminated if a suitably accurate match ($\le 1\textrm{px}$ error) is not found or matching is ambiguous. Due to the small inter-frame displacement, the tracking process is very robust: the largest tolerated prediction error is approximately half a particle diameter. New tracks are initiated by stereo-matching and triangulating unmatched particles. The procedure is iterated until all frames have been processed. Since tracks are frequently interrupted, we reconnect track segments based on their proximity in position-velocity space using the method described in \citet{Xu2008}. 

Due to the finite size of the measurement volume, our measurement of the distribution of particles is sharply truncated near the edges of the measurement volume. This is illustrated in Figure \ref{fig:seeding-density}, which shows isosurfaces of the average seeding density within the measurement volume. To avoid selection biases associated with particles entering and leaving near the edges of the measurement volume, we only sample the statistics of tracers when they are inside the ellipsoidal region shown. The principal diameters of this ellipsoid measure $20.0 \times 27.0 \times 40.8\textrm{mm}^3$. 

\subsection{Direct Numerical Simulation}
\label{sec:datasets:dns}

To provide data for our simulation of our laboratory particle tracking experiment, we ran a  direct numerical simulation of tracer particles in forced, homogeneous, isotropic turbulence at $R_\lambda = 190$ in a $1344^3$ periodic cubic box of side length  $2\pi$.
A standard pseudo-spectral scheme was used to solve the Navier-Stokes equations in their vorticity formulation with statistical stationarity maintained by means of a large-scale band-passed Lundgren forcing in the wavenumber range [1.5,3]. 
This resulted in an integral length-scale over $11$ times smaller than the box size, which helps to minimise effects of the periodic boundary conditions on flow statistics.
The high spatial resolution ($k_{\textrm{max}}\eta=2$, where $k_{\textrm{max}}$ is the maximum resolvable wavenumber) ensures that the small-scales are adequately resolved to capture extreme events.

After achieving statistical stationarity, $10^7$ tracer particles were introduced and advected with the flow and their position, velocity and acceleration were recorded over 0.5 integral time-scales ($10.4\tau_\eta$). 
We stored the tracer state at $\sim0.01\tau_\eta$ intervals in order to be able to recover the extreme acceleration events from the position signal. 
Additional care was taken over the integration of the tracers: cubic splines were used to interpolate the underlying velocity fields and time-stepping was performed using a fourth-order Adams-Bashforth method. 
This effort has ensured that the Lagrangian velocity and acceleration statistics, as obtained from finite differences of trajectories, are in good agreement with those sampled from the fields. 
Further details of the solver can be found in \citet{LalescuWilczek2018}. 

\subsection{Synthetic Particle Tracking Experiment}
\label{sec:datasets:syn}

\begin{figure*}[t]
\includegraphics[width=\textwidth]{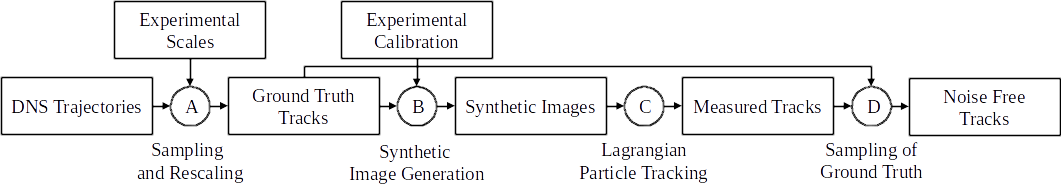}
\caption{Data flow of synthetic particle tracking procedure}
\label{fig:simulation-flow}
\end{figure*}

Trajectories from the DNS were used to simulate the experimental acquisition of particle tracks. The data flow is outlined in Figure \ref{fig:simulation-flow}. There are four steps, which we now outline in detail.

\begin{figure}[!h]
\begin{subfigure}{0.2\textwidth}
	\includegraphics[trim=0cm 0cm 0cm 0cm,height=4.5cm]{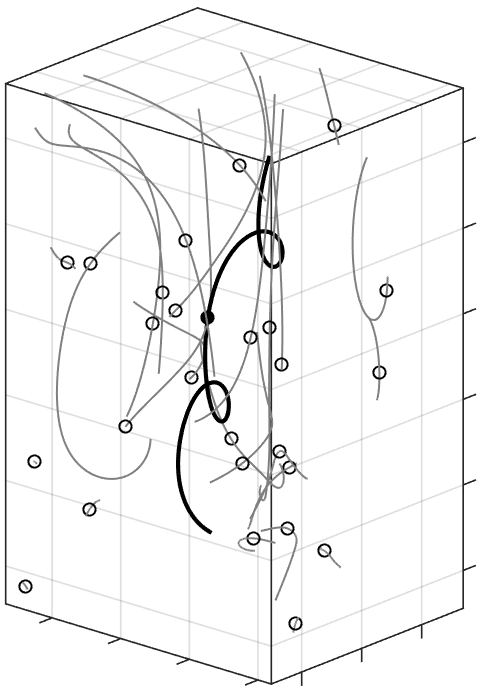}
	\caption{3D Projection}
	\label{fig:track-simulation-3d}
\end{subfigure}
\begin{subfigure}{0.28\textwidth}
	\includegraphics[trim=0cm 0cm 0cm 0cm,height=4.5cm]{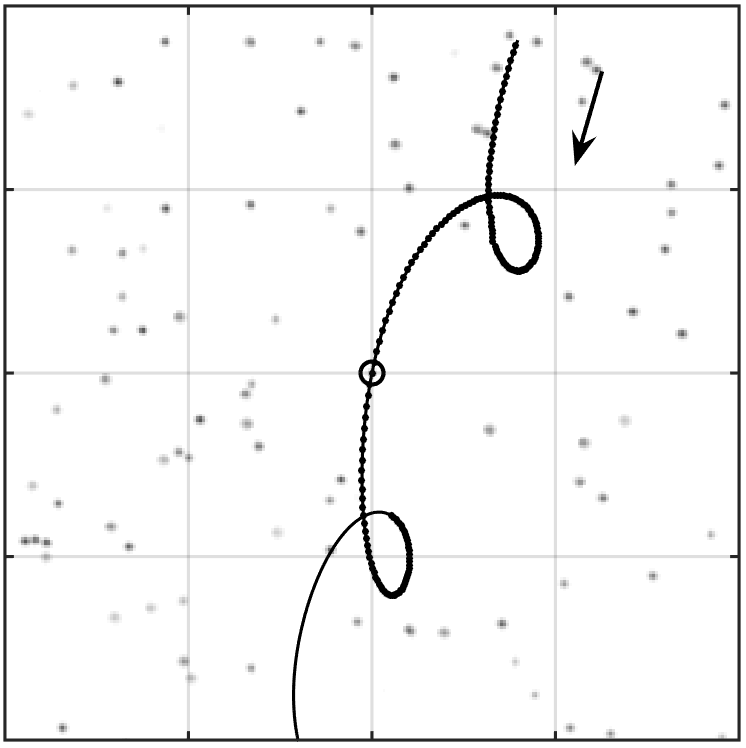}
	\caption{2D Synthetic Image}
	\label{fig:track-simulation-2d}
\end{subfigure}
\caption{Snapshot of a simulated particle, its neighbours and their trajectories. (\subref{fig:track-simulation-3d}) 3D projection of ground truth particle track (bold, filled) and its neighbours (open, grey). The seeding density and Kolmogorov scales are comparable to experiment. (\subref{fig:track-simulation-2d}) Corresponding synthetic particle image for camera 3. The large circle indicates the position at the reference time, whilst small dots indicate the image position recovered from LPT at every fourth timestep. The measured track is incomplete because the particle leaves the measurement volume. Grid markings in both figures correspond to 64 pixels ($2.1\textrm{mm}$).}
\end{figure}

The first step (A) is to sample subsets of the DNS trajectories (corresponding to different sub-volumes from the full simulation) and rescale these from code units to physical units to match the Kolmogorov scales and seeding density of the experiment (see Table \ref{tbl:datasets}). 
This defines the ``ground truth" of the synthetic measurement, a sample of which is shown in Figure \ref{fig:track-simulation-3d}. 
The second step (B) is to generate particle tracking images based on these ground truth data. For this, we projected the positions of tracers in the experimental geometry onto images using same geometric camera calibration used in the experiment.
A procedure similar to the reprojection step of the Shake The Box algorithm was used to render particle images \citep{Schanz2016}. 
These are simulated using three-parameter Gaussian intensity profile, which was obtained from the experimental calibration of the optical transfer function, following \citet{Schanz2013}. Images are then quantised to 8-bit resolution and saved using the same sparse compression format as the experimental data. 
The procedure effectively simulates an experimental measurement of tracer particles from DNS. A sample synthetic particle image is shown in Figure \ref{fig:track-simulation-2d}. 
In total, we generated $5.17\times 10^6$ such images, corresponding to $5040$ subsets of tracers over $1025$ timesteps. 

Using a procedure similar to \citep{Voth2002}, we then processed these synthetic particle images with the same particle tracking and post-processing toolchain as experimental data described in \S\ref{sec:datasets:experiment} (step C). 
This produces an experimental sampling of the ground truth dataset, as illustrated in Figure \ref{fig:track-simulation-2d}. 
The synthetic dataset is matched to the $R_\lambda = 203$ experiment in terms of Reynolds number, scale, measurement geometry, optical properties and processing scheme. 
Amongst others, our modeling simplifications overlook effects such as particle inertia, collisions, polydisperse particle sizing, illumination instability, shadowing, image noise and multiple scattering. These effects are expected to impair measurement quality. 
As such, the synthetic datasets should be considered as a ``best case scenario" measurement analogous to our experimental cases, which incorporate the major sources of bias error. Crucially, the synthetic dataset captures the bias effects we identify in \S\ref{sec:introduction}. 
The availability of ground-truth information permits a quantitative test of the correction methods we describe in \S\ref{sec:methodology}. 

For the simulated measurement, we have sets of tracks representing the ground truth and measurement. The measurement represents a noisy sub-sampling of the ground truth, since some particle trajectories are incompletely registered by the algorithm. In addition, the measurement contains ghost tracks, which do not closely correspond to any track in the ground truth set. To disentangle sampling and noise effects, we constructed a ``noise-free" dataset (step D). This dataset was constructed by identifying all correspondences between the measurement and ground-truth tracks which were sufficiently close (within 1 pixel of position error over their entire lifetime). For this set of ``real" tracks, we replaced the measured position with its ground truth value. Thus, the noise-free dataset represents the experimental sampling of the ground truth without noise.

\section{Simulation Results}
\label{sec:results}

In this section, we apply the error correction techniques described in \S\ref{sec:methodology} to the numerical simulations of Lagrangian Particle Tracking described in \S\ref{sec:datasets:syn}. These are used to validate the use of independent measurements and spline filtering in reducing biases in acceleration moments and probability density functions due to filtering and noise.

\subsection{Acceleration Moments}
\label{sec:results:acceleration}

\begin{figure*}[ht]
\centering
\begin{subfigure}[h]{0.49\textwidth}
	\centering
	\includegraphics[height=0.68\textwidth]{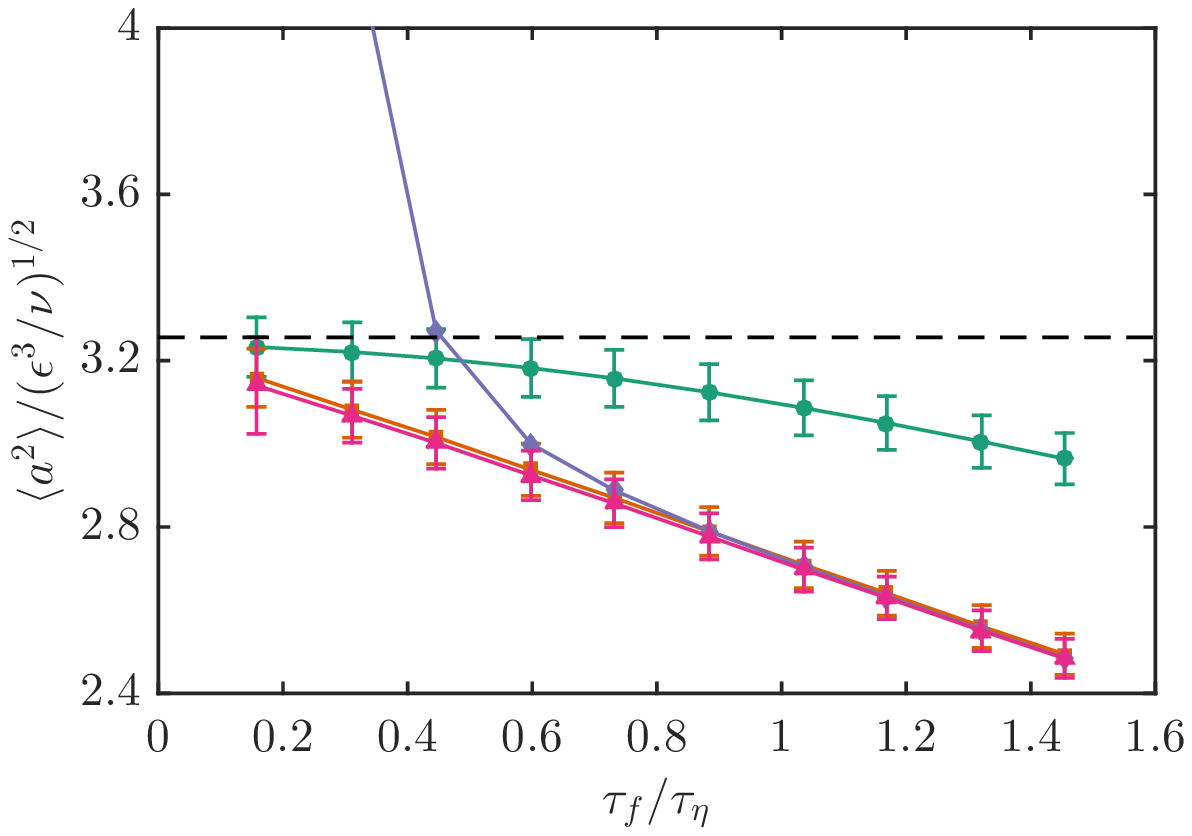}
	\caption{Acceleration variance, Gaussian kernel}
	\label{fig:acc-var-gauss}
\end{subfigure}
\begin{subfigure}[h]{0.49\textwidth}
	\centering
	\includegraphics[height=0.68\textwidth]{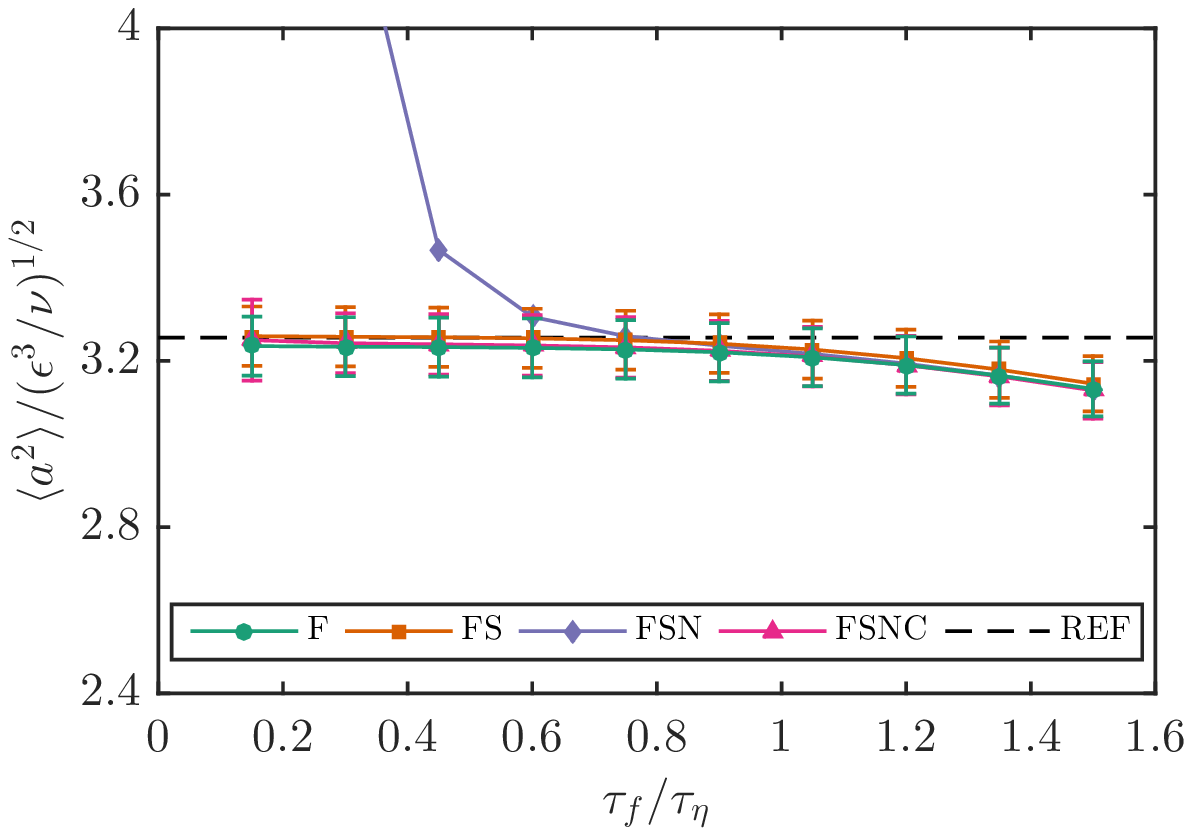}
	\caption{Acceleration variance, penalised spline}
	\label{fig:acc-var-spl}
\end{subfigure}
\begin{subfigure}[h]{0.49\textwidth}
	\centering
	\includegraphics[height=0.68\textwidth]{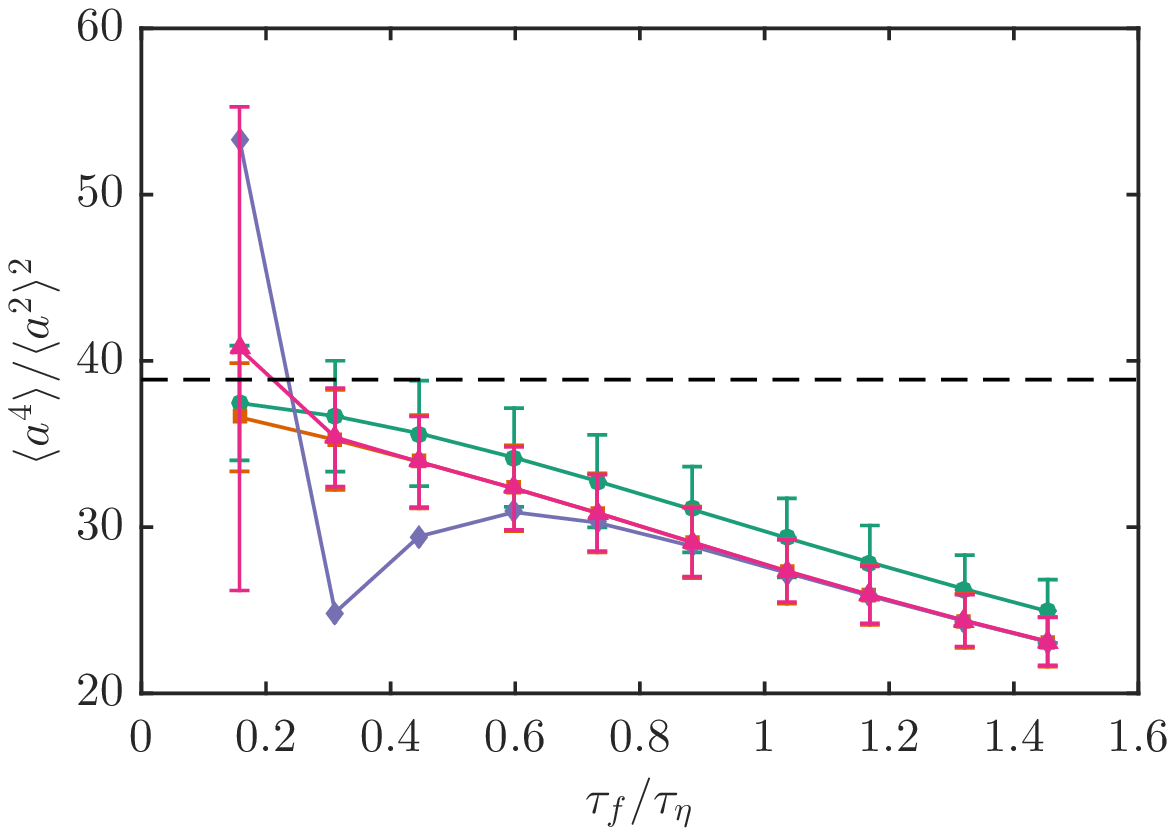}
	\caption{Acceleration flatness, Gaussian kernel}
	\label{fig:acc-flat-gauss}
\end{subfigure}
\begin{subfigure}[h]{0.49\textwidth}
	\centering
	\includegraphics[height=0.68\textwidth]{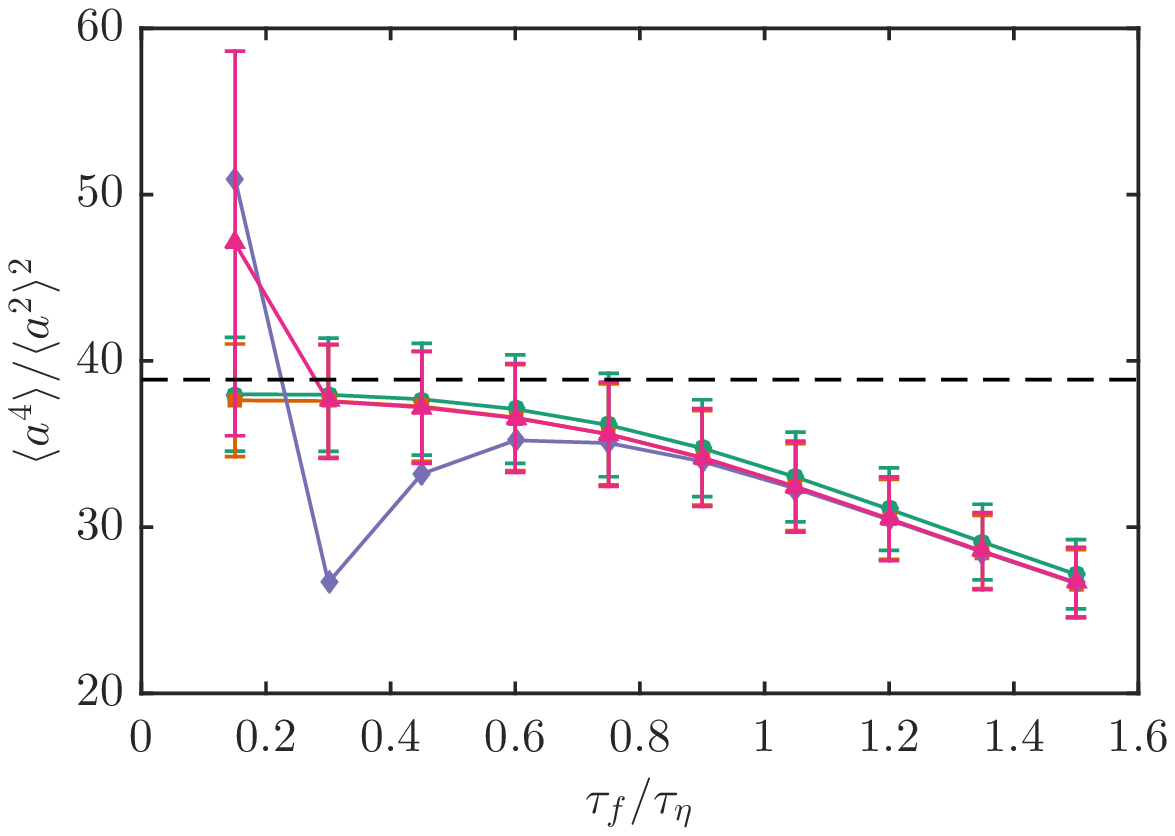}
	\caption{Acceleration flatness, penalised spline}
	\label{fig:acc-flat-spl}
\end{subfigure}
\caption{Acceleration variance (\subref{fig:acc-var-gauss},\subref{fig:acc-var-spl}) and  flatness (\subref{fig:acc-flat-gauss},\subref{fig:acc-flat-spl}) as a function of filter scale and type: Gaussian (\subref{fig:acc-var-gauss},\subref{fig:acc-flat-gauss}) and penalised spline (\subref{fig:acc-var-spl},\subref{fig:acc-flat-spl}). Black dashed line (\sampleline{dashed}) shows the reference value evaluated from DNS. Markers indicate different combinations of bias effects: $\green\medblackcircle$ filtering (F) $\orange\medblacksquare$ filtering and sampling (FS) $\blue\medblackdiamond$ filtering, sampling and noise (FSN) $\magenta\medblacktriangleup$ filtering, sampling, noise and correction (FSNC). Comparisons between curves allow different bias effects to be isolated.  Error bars show bootstrapped 95\% statistical confidence intervals.} 
\label{fig:acc-moments}
\end{figure*}

The statistics of Lagrangian acceleration are remarkably sensitive to measurement noise, filter scale and sampling biases. 
Figures \ref{fig:acc-var-gauss} and \ref{fig:acc-var-spl} show the dependence of the measured acceleration variance upon the filter time-scale $\tau_f = 1/f_{\textrm{ENBW}}$ for the Gaussian and spline filters, respectively.
The variance is calculated for the ground truth, noise-free and simulated measurement datasets.
Additionally, we plot the acceleration variance for the measurement as corrected by (\ref{eqn:a-moments-list}). 
The reference value for the acceleration variance, as obtained from the ground-truth acceleration sampled on the particles, is indicated with the dashed line. 
We note the mean acceleration is negligible in all cases. 

By comparing different curves, we can isolate the effect of different error sources (filtering, selection bias and noise). The effect of filtering is isolated by comparing the two filtered, ground-truth datasets to the reference value. The ensemble in these three cases is identical: only the manner of filtering differs. Both spline and Gaussian filters have a modest effect ($\le 5\%$) upon acceleration variance for a filter scale below $1\tau_\eta$. The Gaussian filter attenuates the signal more than the spline: this may be attributed to the sharper spectral cutoff of the spline filter.

A more pronounced difference between the two filters is seen when examining the noise-free datasets (shown in green). Here, the Gaussian filter exhibits a strong dependence on scale, whilst the spline filtered data does not. Since the Gaussian filter implementation rejects data near track ends and interpolated locations, the sampled ensemble is reduced as the filter support is increased. This introduces a sampling bias by under-representing faster particles, which tend to have shorter tracks and are correlated with larger accelerations \citep{Sawford2003,Voth1998}. Omission of these data results in the underestimation of the acceleration variance.

Unsurprisingly, measurement noise significantly increases the measured acceleration variance when small filter scales are used, as evidenced by the purple curves for the synthetic measurement dataset. 
At the smallest filter scales, the noise power exceeds the signal by a factor of ten. 
Due to the noise, the uncorrected measurement shows no range where the result is insensitive to the choice of scale. 
When noise is accounted for by applying the correction in (\ref{eqn:a-moments-list}) there is remarkable agreement (within $0.6\%$) between the corrected data (magenta) and noise-free data (green). 
The price paid for noise correction is statistical confidence: as the signal to noise ratio is reduced by decreasing the filter scale, the sampling error increases.

The combined effects of filtering, sampling and noise are more significant when considering the higher-order moments of acceleration. 
Figures \ref{fig:acc-flat-gauss} and \ref{fig:acc-flat-spl} show the filter and scale dependence of the acceleration flatness for the Gaussian and spline filters, respectively.
Examination of the filtered ground-truth dataset shows that the acceleration flatness has a stronger dependence on filter scale when a Gaussian filter is used. 
Similarly, the noise-free case shows a much stronger dependence upon filter scale when a Gaussian filter is used in comparison to the smoothing spline. 

The introduction of measurement noise significantly affects the measured flatness. At small filter scales, noise dominates the signal and we effectively measure the flatness factor of the noise. At larger filter scales, the filtered acceleration signal dominates, but the filtering has a strong effect on the signal's flatness factor. Thus, with uncorrected data, no range of filter scales is observed where the result is invariant to scale. Remarkably, when the noise correction (\ref{eqn:a-moments-list}) is applied, excellent agreement (well within statistical confidence) with the noise-free case is observed. Importantly, we observe a range of filter scales where the flatness has only a weak dependence on scale. This means it is possible to conclude, on the basis of the data, whether the filter scale is sufficiently small.

At this stage, two conclusions may be drawn. Firstly, it is most preferable to use cubic smoothing splines for filtering LPT data, as they introduce less sampling bias and have an advantageously sharp spectral cutoff which filters less signal. Secondly, the noise compensation technique for moments described in \S\ref{sec:methodology:noise} has been validated. 

\subsection{Acceleration PDF}

\begin{figure}[t]
\centering
\includegraphics[width=0.48\textwidth]{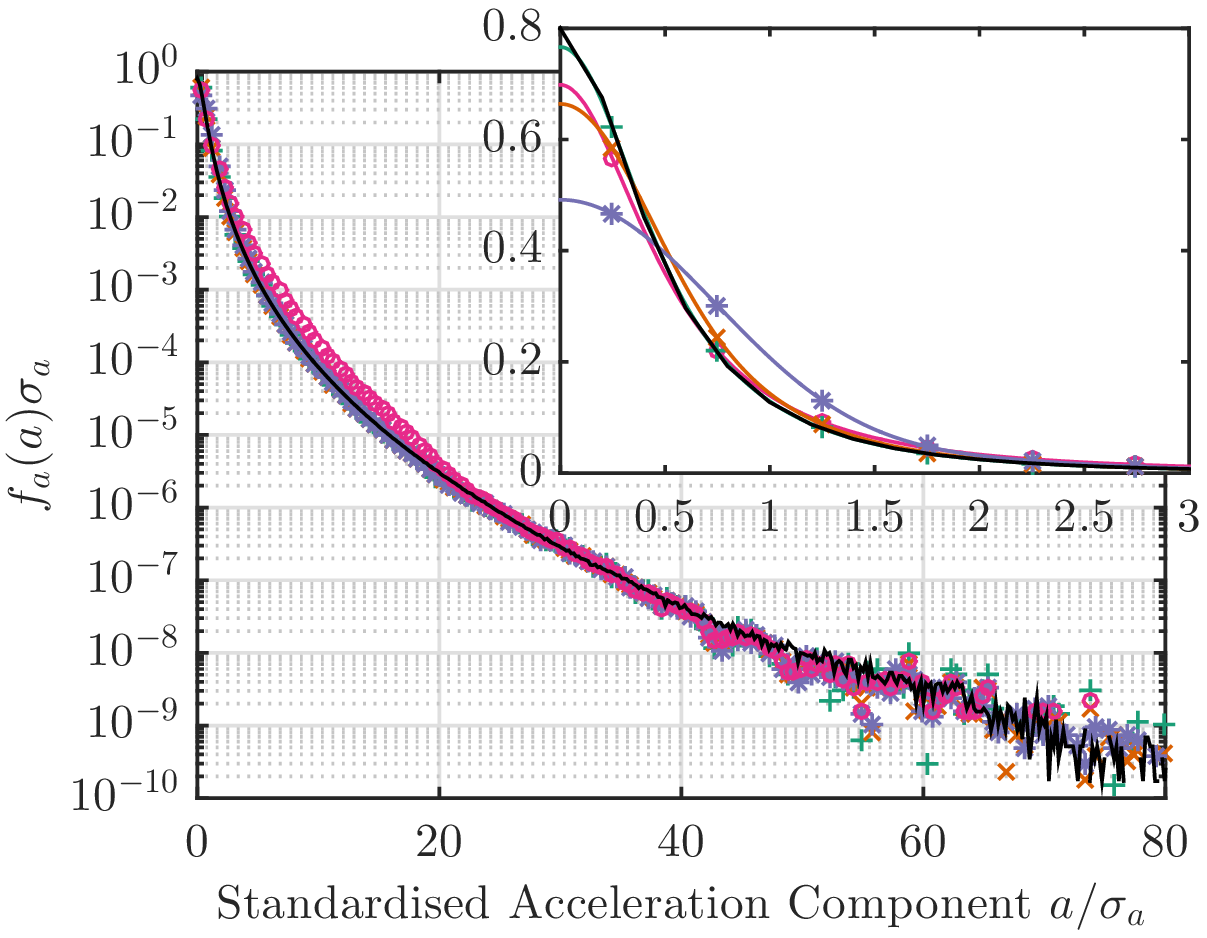}
\caption{
Estimation and recovery of ground truth acceleration PDF using very noisy data. 
The solid, black line (\sampleline{solid}) is the PDF $f_a$ obtained from the DNS reference.
Markers show: $({\magenta\medcircle})$ the measured PDF $f_{\overline{a}}$ without correction, $({\green +},{\orange\times},{\blue *})$ the PDF $\widetilde{f}_a$ with correction at kernel bandwidths $h/\sigma_a = 0.05, 0.10, 0.20$ respectively.
Inset: the same quantities on a linear scale. 
Since the PDFs are symmetric, only the positive half is shown.}
\label{fig:acc-pdf-p1344}
\end{figure}

We now consider the measurement of the acceleration PDF when data is contaminated with noise. 
To demonstrate this, we consider the acceleration distribution obtained from the simulated measurement using a smoothing spline filter with $\tau_f=0.27\tau_\eta$, which yields a signal to noise ratio of only $4\textrm{dB}$.
Figure \ref{fig:acc-pdf-p1344} shows the standardised PDF of acceleration (with and without correction) and directly from the DNS acceleration field (the reference).
The presence of measurement noise overestimates the core of the PDF, but does not significantly influence the tails, which are in close agreement with the reference PDF. 
The deconvolution procedure provides a remarkably accurate correction to the measured PDF, as evidenced by the good agreement with the reference data over a wide range of kernel bandwidths. 

The inset of Figure \ref{fig:acc-pdf-p1344} illustrates the tradeoff which is made when selecting the kernel bandwidth $h$.
Using a smaller bandwidth provides a more accurate estimation of the core, where the underlying data is dense, but undersmooths the tails of the PDF, whereas a larger bandwidth oversmooths the core but captures the tails more accurately.
In general, an ``optimal" bandwidth is difficult to define, since this depends upon the relative importance of statistical uncertainty and accuracy, which will depend the chosen analysis.
We refer the reader to \citet{WangWang2011} for a practical review on the available methods for bandwidth selection.

One way of assessing the accuracy of the corrected PDF is to examine its moments.
Based on the moment-generating properties of the characteristic function, one can show that for the kernel in (\ref{eqn:phi_K}), the second and fourth moments of the corrected PDF $\widetilde{f}_a$ are given by
\begin{equation}
\label{eqn:acc-moments-deconv}
\int \widetilde{f}_a(a) a^2 \mathrm{d}a
= \langle \overline{a}^2 - \widehat{a}^2 \rangle + 6h^2 
\end{equation}
and
\begin{equation}
\int \widetilde{f}_a(a) a^4 \mathrm{d}a
= \langle \overline{a}^4 - \widehat{a}^4 \rangle - 6\langle \overline{a}^2  - \widehat{a}^2\rangle(\langle \widehat{a}^2 \rangle + 6h^2) + 72h^4
\end{equation}
respectively. For the above choice of bandwidth and noise level, this gives a relative error of $6\%$ in the second moment and $0.94\%$ in the fourth moment, which are comparable to the sampling error.

\section{Experimental Results}
\label{sec:exp-results}

In this section, we evaluate the statistics of Lagrangian acceleration for the experimental datasets described in \S\ref{sec:datasets:experiment} and make comparisons to the wider literature. This provides a direct test of the bias correction techniques outlined in \S\ref{sec:methodology} over a wide range of Reynolds numbers ($R_\lambda = 109 - 504$), spatial resolution ($\eta = 1.3 - 11\textrm{px}$) and effective temporal resolution ($\tau_f/\tau_\eta = 0.15 - 1.5$).

\subsection{Acceleration Moments}
\label{sec:exp-results:acceleration}

\begin{figure}[h!]
\centering
\begin{subfigure}{0.47\textwidth}
	\includegraphics[trim=-0.4cm 0cm 0cm 0cm, clip, width=\textwidth]{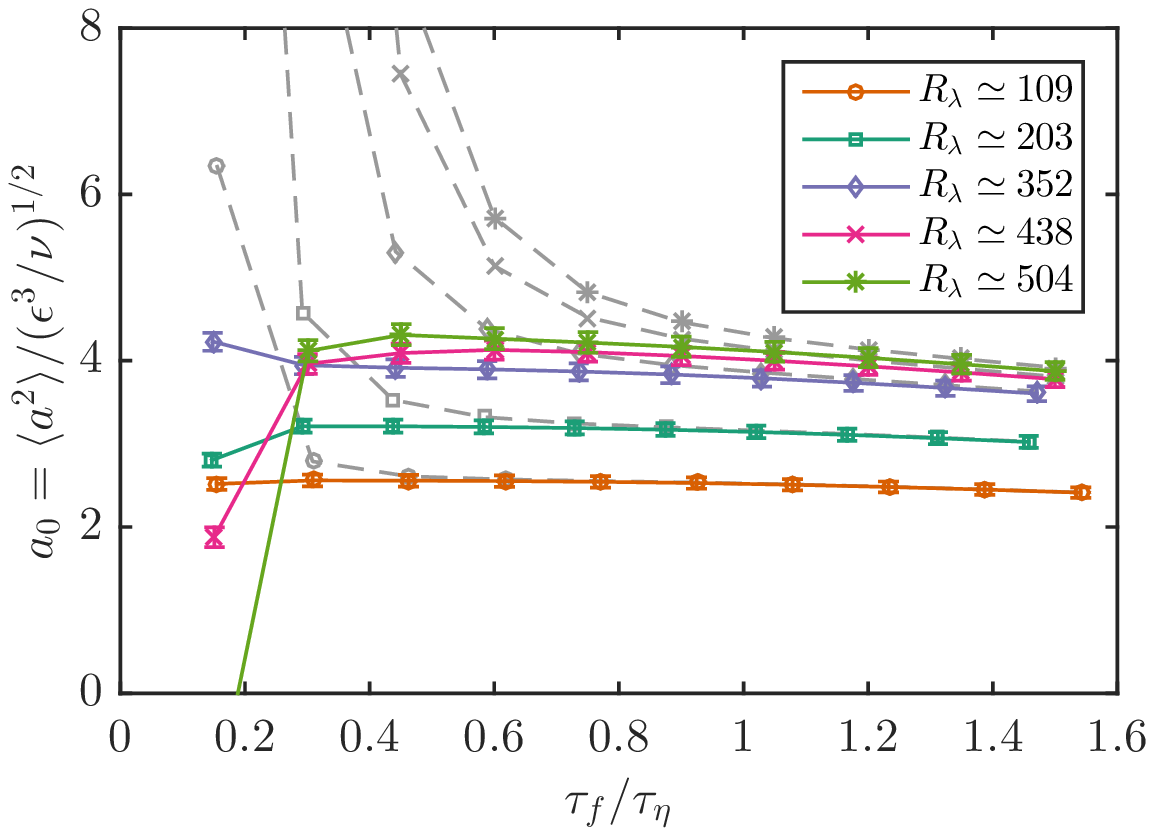}
	\caption{Variance}
	\label{fig:acc-var-exp}
\end{subfigure}
\begin{subfigure}{0.47\textwidth}
	\includegraphics[trim=0cm 0cm 0cm 0cm, clip, width=\textwidth]{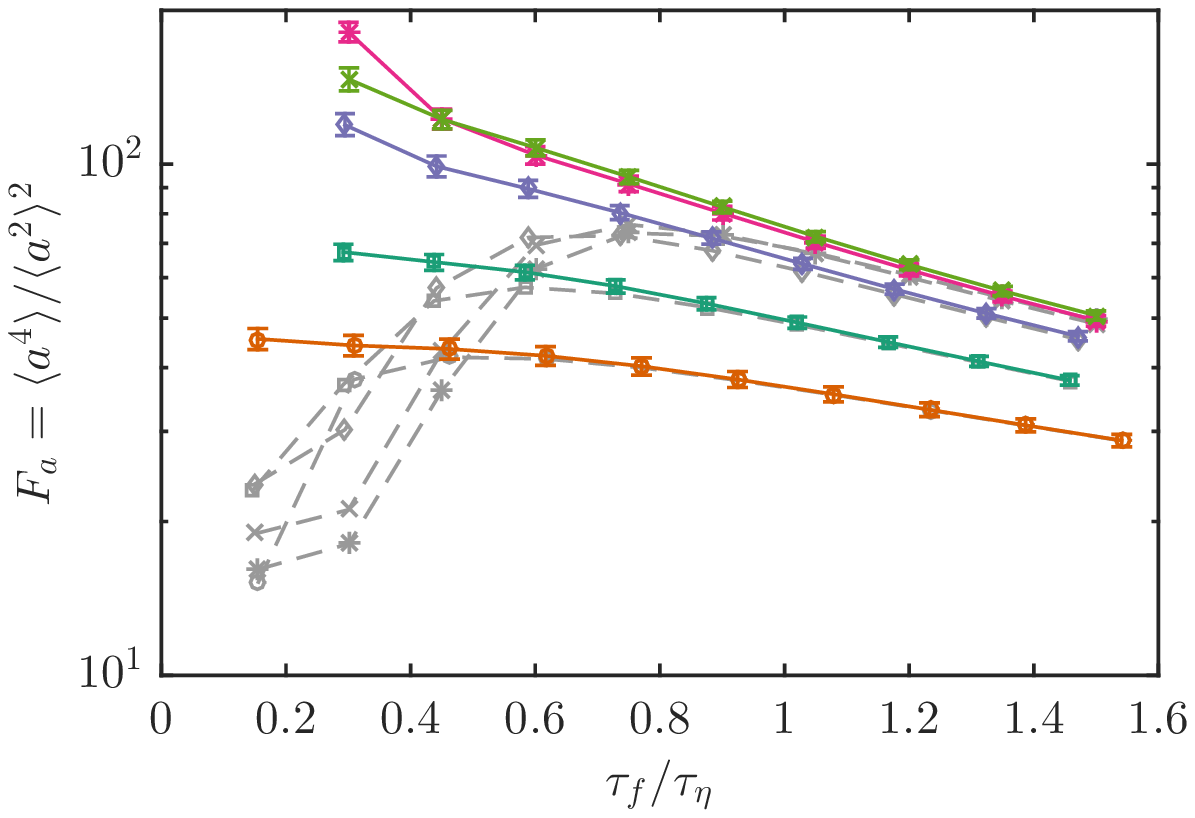} 
	\caption{Flatness}
	\label{fig:acc-flat-exp}
\end{subfigure}
\label{fig:acc-moments-exp}
\caption{Experimental measurement of acceleration (\subref{fig:acc-var-exp}) variance and (\subref{fig:acc-flat-exp}) flatness factor as a function of filter scale. Solid lines (\sampleline{solid}) show corrected moments, dashed lines (\sampleline{dashed}) show uncorrected moments. Markers $\medcircle, \medsquare, \meddiamond, \times, \ast$ show Reynolds number in ascending order. Error bars show 90\% confidence intervals.}
\end{figure}

Figure \ref{fig:acc-var-exp} shows measurements of the acceleration variance as a function of filter scale for our experimental datasets. 
A penalised spline filter was used. 
The directional dependence of this measurement is negligible ($\le 3.1\%$ variation across cameras). 
Both the raw (dashed line) and noise corrected (solid line) measurement are shown. It is readily apparent that, as the filter scale is reduced, the noise contribution dominates the measurement of acceleration variance. 
However, when this noise contribution is corrected for, we see that the acceleration has a weak dependence on the filter scale, which is the expected physical behaviour.
When the noise level is made very large, the correction technique starts to break down.
This is because, in practice, there exists a small correlation in the measurement error between cameras which becomes significant when the noise level is very large (in our data, below $\textrm{SNR} \sim 0\textrm{dB}$). 
This demonstrates that the noise-correction technique is able to accurately compensate second moments of moderately noisy experimental data.

It is worth emphasising that, when penalised spline filtering is used in conjunction with noise correction to estimate the acceleration variance, the result has a very weak dependence on filter scale.
For example, at $R_\lambda = 509$ where the scale dependence is strongest, doubling the filter bandwidth from $\tau_f=0.9\tau_\eta$ to $\tau_f=0.45\tau_\eta$ (i.e. from $f_{\textrm{ENBW}}\tau_\eta = 1.1$ to $2.2$) increases the estimated variance by less than $3.5\%$, which is comparable to the statistical uncertainty. 
This is in stark contrast to the exponential dependence upon filter scale observed by \citet{Voth2002} and \citet{Crawford2004}.
Moreover, the weak scale dependence allows a quantitative assessment of whether the filter scale is sufficiently small.

Corresponding measurements of the acceleration flatness are shown in Figure \ref{fig:acc-flat-exp}. 
As the filter scale is reduced, the uncorrected acceleration flatness first increases as finer scale flow features are probed, then decreases as the noise contribution begins to dominate. 
By correcting for noise, we see a steady increase in the flatness factor as the filter scale is reduced. 
The presence of this expected physical behaviour qualitatively confirms the validity of the noise correction approach.
In contrast to the numerical simulations presented in Figure \ref{fig:acc-flat-spl}, we do not have sufficient resolution to identify clear plateau region at all but the lowest Reynolds numbers. 
This indicates that the experimental acceleration distribution contains contributions from very rapid motions not present in the simulations.
We speculate that the difference arises due to the intermittency of the large-scale forcing in the LEM, which may occasionally generate regions of intense turbulence where the fastest dynamics proceed on timescales below the average Kolmogorov scale.

\begin{figure}
\includegraphics[trim=0cm 0cm 0cm 0cm, clip, width=0.48\textwidth]{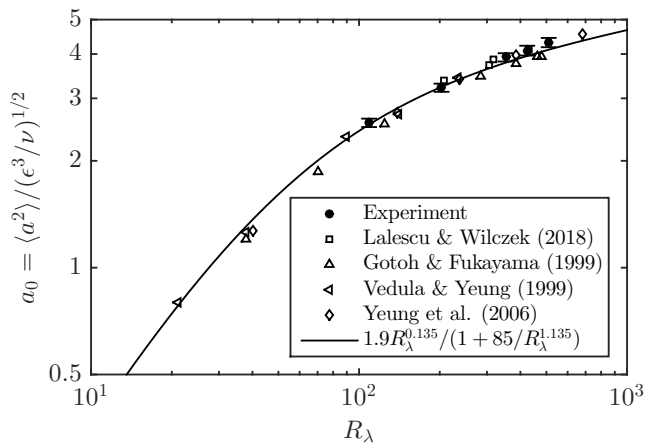}
\caption{Collapse of normalised acceleration variance $a_0$ obtained from present experimental data (for $\tau_f = 0.6\tau_\eta$) with data from simulations of homogeneous, isotropic turbulence and multifractal model fit \citep{Sawford2003}, plotted as a function of $R_\lambda$. Error bars show 90\% confidence intervals.}
\label{fig:a0-comparison}
\end{figure}

We now compare our experimental measurements of the acceleration variance to measurements and models reported in the literature. 
Figure \ref{fig:a0-comparison} shows the Heisenberg-Yaglom coefficient $a_0  = \langle a^2\rangle/(\epsilon^3/\nu)^{1/2}$ obtained from the present experiments, simulations of homogeneous, isotropic turbulence \citep{VedulaYeung1999,GotohFukayama2001,Yeung2006,LalescuWilczek2018} and the multifractal model fit obtained by \citet{Sawford2003}.
Note that the multifractal model was obtained by \citep{Sawford2003} by a fit to the data from \citet{VedulaYeung1999} and \citet{GotohFukayama2001}.
The experimental data is in reasonable agreement with the empirical fit and show a comparable degree of scatter to the available numerical data. 
This quantitative agreement demonstrates the ability of the experimental technique to eliminate systematic biases when measuring the acceleration variance.

\subsection{Acceleration PDF}
\label{sec:exp-results:acceleration-pdf}

\begin{figure}
\includegraphics[trim=0cm 0cm 0cm -0cm, clip, width=0.48\textwidth]{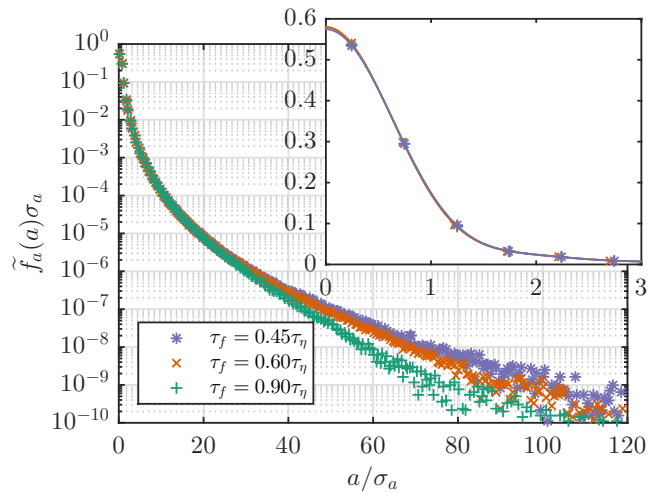} 
\caption{Self-consistency of the corrected, experimental distribution of acceleration $\widetilde{f}_a$ at $R_\lambda = 504$, as a function of filter scale. 
The kernel bandwidth is $h = 0.2\sigma_a$.
Inset: the same distribution on a linear scale.
For clarity, only the positive half of the PDF is shown.}
\label{fig:acc-pdf-iso960}
\end{figure}

Unlike the simulation experiment, there is no ground truth available to compare the measured acceleration distribution to validate the measurement. 
However, we are able to test the consistency of the deconvolved PDF as a function of filter scale: when the filter scale is sufficiently small, only the tails of the distribution should change significantly.
This is demonstrated in Figure \ref{fig:acc-pdf-iso960}, which shows the corrected distribution of acceleration $\widetilde{f}_a$ at filter scales $\tau_f = 0.45 - 0.9\tau_\eta$.
As the filter scale is increased, the core of the distribition remains largely unchanged, whereas the probability density in the far tails increases markedly.

It is remarkable that, even though the signal to noise ratio is $0\textrm{dB}$ for $\tau_f=0.45\tau_\eta$, the cores of the corrected PDFs are in good agreement.
In this case, deconvolution allows us to improve the temporal resolution of the measurement by a factor of approximately two.
To do the same by reducing the noise level would require a fourfold improvement in spatial resolution.
Implementing deconvolution is, of course, substantially simpler.

\section{Conclusion}
\label{sec:conclusion}

We have presented methods to mitigate two key sources of bias error in LPT measurements, namely noise and filtering effects.
The methods have been validated through the use of numerical simulations of LPT and demonstrated to work in practice via application to experimental LPT measurements in homogeneous, isotropic turbulence over a wide range of Reynolds numbers and effective spatial and temporal resolutions. 

We have outlined methods to correct statistical moments and probability distributions of Lagrangian quantities obtained from LPT data contaminated with noise.
The operating principle is to obtain a measurement of the noise distribution imposed on the velocity or acceleration signal using simultaneous measurements made across two or more cameras. 
This distribution can be used to correct the desired statistics. 

The noise correction technique is mainly limited by the increase in statistical uncertainty associated with the noise correction and the requirement that the noise remain statistically independent of the signal. 
The technique has potential for very general application. 
Examples include the measurement of velocity and Reynolds stress profiles in the near wall boundary layer, which are notoriously difficult to measure accurately \citep{Kaehler2012}.

The use of finite impulse response filters to measure particle acceleration can introduce a significant, systematic error in acceleration statistics. 
Such filters only sample the signal where there is sufficient support, which under-represents shorter, faster tracks correlated with larger accelerations. 
This bias is responsible for introducing a strong dependence of the measured acceleration statistics upon filter scale. 
Instead, we advocate the use of penalised, cubic splines to implement the numerical differentiation and filtering of tracks. 
When spline filtering is used, acceleration statistics are seen to display a much weaker dependence upon filter scale. 
This is because the acceleration may be sampled along the entire length of each track. 

When spline filtering is used in combination with noise correction, a range of filter scales are observed where the acceleration statistics depend only weakly on the filter scale. 
This has not been achieved in the literature to date. 
Crucially, this allows the experimentalist to determine quantitatively whether the chosen filter bandwidth is sufficiently small.
In our measurements of acceleration variance, we find that a filter bandwidth $f_{\textrm{ENBW}}\tau_\eta \ge 1.1$ is sufficient to recover $\ge 96.5\%$ of the signal energy. 
The acceleration flatness requires more bandwidth, depending on the Reynolds number.
The simultaneous application of these techniques allows a substantial reduction in the temporal and spatial resolution required to make accurate measurements of acceleration distributions in turbulent flows and Lagrangian properties of turbulent flows in general.

The generality and simplicity of the above techniques allows us to suggest the adoption of the following best-practices in making Lagrangian acceleration measurements. 
Discrete convolution methods for numerical differentiation should be avoided due to the sampling errors they introduce by under-representing data near ends of tracks and in interpolated segments. 
Filter time-scales should be chosen in a range where they do not influence the results.
Systematic errors in statistical moments and probability distributions should be quantified or corrected for via the use of simultaneous, independent measurements across independent cameras where possible.

\begin{acknowledgements}
The authors gratefully acknowledge the financial support of the Max Planck Society and EuHIT: European High-Performance Infrastructures in Turbulence, which is funded by the European Commission Framework Program 7 (Grant No. 312778). 
Part of the computations were performed on the clusters of the Max Planck Computing and Data Facility. 
\end{acknowledgements}

\FloatBarrier
\bibliographystyle{spbasic}      
\bibliography{bias}   

\end{document}